\documentclass[12pt]{iopart}
\usepackage{graphicx} 
\usepackage{bm}
\usepackage{mathrsfs}

\usepackage{lmodern,amssymb,amsthm}
\usepackage{iopams}
\usepackage{xspace}
\usepackage{hyperref}
\usepackage{graphicx}
\usepackage{calc}
\usepackage[utf8]{inputenc}
\usepackage[T1]{fontenc}
\usepackage{verbatim} 

\newcommand{\scri}{\ensuremath{\mathscr{I}^+}\xspace}
\newcommand{\sh}{\sigma_{\rm h}}

\newcommand{\beq}{\begin{equation}}
\newcommand{\eeq}{\end{equation}}
\newcommand{\bea}{\begin{eqnarray}}
\newcommand{\eea}{\end{eqnarray}}
\newcommand{\dfrac}[2]{{\displaystyle\frac{#1}{#2}}}

\newcommand{\TPI}{\address{Theoretisch-Physikalisches Institut,
    Friedrich-Schiller-Universität Jena,\\ Max-Wien-Platz 1,
          D-07743 Jena, Germany}}
\begin{document}

\title{Axisymmetric constant mean curvature slices in the Kerr space-time}
\author{David Schinkel, Rodrigo Panosso Macedo, Marcus Ansorg} 
\TPI
\date{\today}

\begin{abstract}
Recently, there have been efforts to solve Einstein's equation in the context of a conformal compactification of space-time. Of particular importance in this regard are the so called CMC-foliations, characterized by spatial hyperboloidal hypersurfaces with a constant {\em extrinsic mean curvature} $K$. However, although of interest for general space-times, CMC-slices are known explicitly only for the spherically symmetric Schwarzschild metric. This work is devoted to numerically determining axisymmetric CMC-slices within the Kerr solution. We construct such slices outside the black hole horizon through an appropriate coordinate transformation in which an unknown auxiliary function $A$ is involved. The condition $K={\rm const}$ throughout the slice leads to a nonlinear partial differential equation for the function $A$, which is solved with a pseudo-spectral method.
The results exhibit exponential convergence, as is to be expected in a pseudo-spectral scheme for analytic solutions.  As a by-product, we identify CMC-slices of the Schwarzschild solution which are not spherically symmetric.
\end{abstract}
\maketitle
\section{Introduction}
The construction of black hole spacetimes in terms of hyperboloidal slices which extend up to future null infinity \scri is widely accepted as a promising concept to avoid artificial spatial boundaries and to permit a detailed analysis of the outgoing radiation \cite{Bondi1962,Stewart1989}. The very fact that gravitational wave signals are well defined at null infinity only has led to a considerable interest in this approach, especially in the realm of numerical computations. The hope is that eventually the advanced extrapolation algorithms to compute the radiation at \scri, which are being used nowadays (e.g.~in the active field of coalescing binary black holes, see \cite{Pfeiffer:2012pc} for a review), can be replaced in favor of direct reading the data at \scri.

In order to include \scri as a finite boundary to the computational domain, it is necessary to {\em compactify} the  hyperboloidal slices (see \cite{Frauendiener:ConformalInfinity} for a review). Yet, the {\em physical metric} becomes singular at \scri, and by splitting it in terms of a {\em conformal factor}, including the singularity, and a regular {\em conformal metric}, we are led to a form of the Einstein equations which is not manifestly regular at \scri. Friedrich \cite{Friedrich:1983} was able to rewrite the system in terms of a symmetric hyperbolic formulation which is equivalent to the Einstein equations and manifestly regular at \scri. In spite of this success and some numerical applications in the weak relativistic regime \cite{Hubner:2000pb}, recent efforts have returned to formulations which consist of considerably smaller systems than the Friedrich one, but have to deal with the irregular terms arising at \scri \cite{Zenginoglu:2008pw,Moncrief2009}. As a proof of principle, Rinne \cite{Rinne:2009qx} was able to demonstrate in an axisymmetric context that such a system can be stably evolved using standard numerical techniques. 

It is interesting to note that most of these systems make use of {\em constant mean curvature slices} ({\em ``CMC-slices''}). In particular, Moncrief and Rinne work on these slices \cite{Moncrief2009, Rinne:2009qx}, and also the new tetrad formalism by Bardeen et al.~\cite{Bardeen:2011ip} (see also \cite{Bardeen:2011pd}) is given thereupon. Consequently, for such schemes {\em initial data} need to be constructed on CMC-slices.

Initial data arise as the solution of the constraint equations on the initial hypersurface. In general, when such data are constructed on hyperboloidal slices, they will suffer from logarithmic singularities at \scri. A way to avoid such undesired behavior was found by Andersson and Chruściel \cite{Andersson:1992yk,Andersson:1993we, Andersson1994} who proved regularity conditions for initial data on CMC-slices. In addition, on CMC-slices the momentum and Hamiltonian constraints decouple. 

Work along this line used the assumption of isotropic extrinsic curvature \cite{Frauendiener:1998ud, Husa:2005ns}. Alternatively, one may impose the conformal flatness condition for which the momentum constraint possesses an analytically known solution \cite{Bowen:1980yu}. In this context, solutions were constructed by Buchman et al.~\cite{Buchman:2009ew} in order to describe single and binary black hole initial data. Although the corresponding algorithms have been used to create data corresponding to strongly spinning black holes, it was argued in  \cite{Lovelace2008} that such data suffer from undesired (``junk'') radiation. Rather, for the description of highly spinning objects, the conformal flatness condition has to be abandoned, as was shown in \cite{Garat:2000pn}. In \cite{Lovelace2008} it was demonstrated that the amount of undesired radiation can be reduced through quasi-equilibrium initial data based on the superposition of Kerr-Schild metrics.

For Schwarzschild black holes, CMC-slices have been discussed analytically in \cite{brill:2789, Malec:2009}. Numerical work to simulate a single spherically-symmetric black hole in CMC-foliations was carried out in \cite{Gentle:2000aq}. At that point, this approach did not impose a proper compactification of \scri, which, however, was performed in  \cite{Zenginoglu:2008wc}. Another occurrence of CMC-slices of the Schwarzschild metric can be found in \cite{Zenginoglu:2008uc} where they are used as a background to study gravitational perturbations. Note that recently CMC-slices of the Reissner-Nordström black hole were identified in \cite{Tuite2013}.

Naturally, the construction of initial data for highly spinning black holes should proceed from the unperturbed Kerr solution. Hyperboloidal foliations of the Kerr space-time were utilized in \cite{Racz:2011qu, Harms:2013ib} where the late-time behavior of scalar fields on a fixed Kerr background was examined. Note that in these works, the coordinates penetrate smoothly through the black hole horizon. The wave equations obeyed by the scalar fields were considered outside the horizon, i.e.~the so-called excision technique was imposed which was studied in \cite{Thornburg:1987} and further developed in \cite{Seidel:1992} (see also references in \cite{baumgarte2010numerical}). Note that excision is not essential for the conformal concept\footnote{In fact, it would be interesting to construct compactified hyperboloidal slices covering the entirety of the Kerr space time.}, but for most numerical black-hole computations, it suffices to consider the black hole exterior.

In this paper, we provide axisymmetric hyperboloidal CMC-slices of the Kerr metric outside the black hole horizon in horizon-penetrating coordinates. Once given  such foliations one can, in a future project, calculate {\em perturbed} Kerr data on CMC-slices utilizing e.g.~the methods in \cite{Schinkel:2013zm}. Now, as mentioned above, perturbed Kerr data on hyperboloidal CMC-slices would be interesting initial data for the time evolution scheme by Moncrief and Rinne  as well as for the tetrad formalism by Bardeen et al. Also, only in the context of CMC-slices it was shown mathematically that logarithmic singularities at \scri can be avoided in the initial data (although the reasoning and results in \cite{Schinkel:2013zm} support the conjecture that the weaker condition of merely {\em asymptotically} constant mean curvature would suffice). Furthermore, in a subsequent dynamical evolution by virtue of Friedrich's conformal system, the regularity would be retained. In sum, it is the issue of regularity that justifies the construction of CMC-slices.

The paper is organized as follows. In section \ref{sec:HyperboloidalSlicing} we describe hyperboloidal slices of the Kerr metric. Proceeding from Kerr coordinates $(V,r,\theta,\varphi)$, these slices are constructed by performing a specific coordinate transformation in which an auxiliary function $A$ is involved. In section \ref{sec:ConstructionCMC} we show how the CMC-condition $K={\rm constant}$ can be cast into a nonlinear partial differential equation of second order for $A$. As described in section \ref{sec:NumScheme}, the solution of this equation is obtained by utilizing a pseudo-spectral scheme which yields exponential convergence, thus providing strong evidence for the analyticity of the function $A$. Finally, in section \ref{sec:Num_results} we discuss our numerical findings. As a by-product, we demonstrate that there are
CMC-slices of Schwarzschild black holes which are not spherically symmetric. 

We use the following conventions. Greek indices run from 0 to 3 where 0 denotes the time component. Small Latin indices are from 1 to 3, covering spatial indices. Partial derivatives are denoted by a comma and conformal (unphysical) objects will be decorated with a tilde. We use units in which the speed of light as well as Newton's constant of gravitation are unity.
\section{Hyperboloidal slices in the Kerr space-time}\label{sec:HyperboloidalSlicing}

For the transition to compactified hyperboloidal slices in the Kerr metric with mass $M$ and angular momentum $J=aM$, we proceed as in \cite{Schinkel:2013zm}. Starting from Kerr coordinates ($V, r,\theta, \varphi$), in which the line element is horizon-penetrating,
\bea
 ds^2=-\left(1-\frac{2M r}{\rho ^2}\right)dV^2 + 2dV dr -\frac{4M r a}{\rho ^2}\sin^2\theta\, dV d\varphi \nonumber \\ + \rho ^2d\theta^2 - 2a \sin^2\theta\,drd\varphi + \frac{1}{\rho ^2}\left(\left(r^2+a^2\right)^2 - \Delta  a^2\sin^2\theta\right)\sin^2\theta\,d\varphi^2.
 \label{eqn:KerrCoordinates}
\eea
with
\beq
 \rho =\sqrt{r^2+a^2\cos^2\theta}, \qquad \Delta =r^2-2M r+a^2,
\eeq
we perform a coordinate transformation that is motivated by an asymptotic integration of outgoing radial null rays and  equivalent to the {\em height function technique} (see e.g. \cite{Zenginoglu:2007jw}),
\begin{eqnarray}
r &=& \frac{2M}{\sigma}\label{eq:CoordTrafo_r}\\
V &=& 4M\Bigg[\tau + \left(\frac{1}{\sigma} - \log(\sigma)  + A(\sigma,\cos\theta) \right)\Bigg]\label{eq:CoordTrafo_V}.
\end{eqnarray}
In the new coordinates $(\tau,\sigma,\theta,\varphi)$, the compactified hyperboloidal slices are described by $\tau={\rm constant}$. The coordinate location of future null infinity $\scri$ is given by $\sigma=0$ while the  event horizon is located at \begin{equation}
\label{eq:sh}
\sigma=\sh=\frac{2}{1+\sqrt{1-j^2}},\qquad\mbox{with}\quad j=\frac{J}{M^2}.
\end{equation}
In the coordinate transformation (\ref{eq:CoordTrafo_V}) we have included an auxiliary function\footnote{In this work, we restrict ourselves to coordinates in which the axial symmetry as well as the stationarity of the Kerr space-time is reflected by the fact that the metric coefficients do not depend on $\varphi$ and $\tau$.}
\[A=A(\sigma,\mu) \qquad\mbox{with}\quad\mu=\cos\theta,\]
which is supposed to be analytic for all 
\[(\sigma,\mu)\in [0,\sh]\times[-1,1].\]
In principle, any such analytic function $A$ can be chosen, and the corresponding slices $\tau={\rm constant}$ will be hyperboloidal and horizon-penetrating. The only restriction is the requirement that these slices be space-like everywhere away from \scri. In \cite{Schinkel:2013zm}, we imposed asymptotic conditions on $A$, in order to realize {\em asymptotically} constant mean curvature slices, but used the remaining freedom to render rapidly converging spectral expansions of a representative quantity (the conformal lapse $\tilde\alpha$). In this paper, we aim at the determination of $A$ such that, in the coordinates $(\tau,\sigma,\theta,\varphi)$, the mean extrinsic curvature $K$ is constant throughout the slice $\tau={\rm constant}$.
\section{The condition of constant mean curvature}\label{sec:ConstructionCMC}

The mean curvature $K$ is defined as the trace of the extrinsic curvature\footnote{Note that we follow the sign convention employed by Wald \cite{Wald:1984} which is opposite to the widely adopted one by Misner, Thorne \& Wheeler \cite{MTW:1973}. In the Wald convention, the mean curvature $K$ assumes {\em positive} values at \scri.}
\beq
	K_{\mu\nu}=\frac{1}{2}\mathcal{L}_ng_{\mu\nu},
\eeq
where $\mathcal{L}_n$ denotes the Lie derivative along the future pointing unit vector $n^\mu$ normal to the slices $\tau={\rm constant}$. It thus follows that
\beq\label{eq:K}
K=\nabla_\mu n^\mu.
\eeq
As mentioned in the introduction, the coefficients of the {\em physical metric} $g_{\mu\nu}$ with respect to the coordinates $(\tau,\sigma,\theta,\varphi)$ diverge at \scri. Concretely, these coefficients can be written in the form
\beq
g_{\mu\nu}=\Omega^{-2}\tilde{g}_{\mu\nu}.
\eeq  
where $\Omega$ is a {\em conformal factor} which vanishes linearly at \scri,
\[\Omega\Big|_{\scri}=0,\qquad  \frac{\partial\Omega}{\partial\sigma}\Big|_{\scri}\ne 0,\]
and $\tilde{g}_{\mu\nu}$ are the components of a regular {\em conformal metric}. In contrast to $g_{\mu\nu}$, the mean curvature $K$ remains finite at \scri.

It is now straightforward to derive $K$ in the coordinates $(\tau,\sigma,\theta,\varphi)$ as an expression which involves the auxiliary function $A$ (and its first and second derivatives with respect to $\sigma$ and $\mu$). In so doing, we found it most convenient to write $K$ in terms of the {\em conformal lapse $\tilde{\alpha}$},
\begin{equation}
	\begin{array}{lll}
		\tilde{\alpha}&=&\Omega\alpha =\left(-\tilde{g}^{00}\right)^{-1/2} \\[4mm]
      					  &=&\dfrac{1}{2}\sqrt{4 + j^2\mu^2\sigma^2}\biggl\{4(1+\sigma) + \dfrac{j^2}{4}[\mu^2-(1+2\sigma)^2] -4(1-\mu^2)A_{,\mu}^2 \\[4mm]
					 &+&[2 j^2 \sigma ^3+(j^2-8) \sigma ^2+4]A_{,\sigma} -\sigma ^2 \left(j^2 \sigma^2-4 \sigma +4\right)A_{,\sigma}^2\biggr\}^{-1/2},
\end{array}\label{eq:conformal_lapse}
\end{equation}
where we chose the conformal factor to be
\beq
\Omega=\frac{\sigma}{4M}.
\eeq
We thus obtain for the mean curvature (\ref{eq:K}) the following expression:
\beq\label{eq:K_explicit}
K=\frac{1}{2M(4+j^2\mu^2\sigma^2)}\sum\limits_{i=0}^{5} p_i \sigma^i,
\eeq
where the coefficients $p_i$ are defined by
\begin{equation}
\label{eq:p_i_coefficients}
	\begin{array}{lll}
		p_0&=&12 \tilde{\alpha}\\[2mm]
p_1&=& -4\tilde\alpha_{,\sigma} + 8 \left[A_{,\mu}\tilde\alpha (1-\mu^2)\right]_{,\mu} \\[2mm]
p_2&=&-\tilde{\alpha} (8 A_{,\sigma}-j^2+8)\\[2mm]
p_3&=&\tilde{\alpha}_{,\sigma} (8 A_{,\sigma}-j^2+8)+8 A_{,\sigma\sigma} \tilde{\alpha}\\[2mm]
p_4&=&2 \tilde{\alpha} (j^2 A_{,\sigma}-4 A_{,\sigma\sigma})-2 \tilde{\alpha}_{,\sigma} (4 A_{,\sigma}+j^2)\\[2mm]
p_5&=&2 j^2 (A_{,\sigma} \tilde{\alpha}_{,\sigma}+A_{,\sigma\sigma} \tilde{\alpha}).
\end{array}
\end{equation}
Equation (\ref{eq:K_explicit}) with the coefficients (\ref{eq:p_i_coefficients}) describes a nonlinear equation for the auxiliary coordinate function $A$. Although this equation is linear in the second order derivatives, it appears difficult to determine its type, since the coefficients in front of the second derivatives contain complicated expressions involving $A$ and its first derivatives. 

We analyse this equation through a Taylor expansion in the vicinity of \scri. Defining
\[A_i(\mu)=\frac{\partial^i A}{\partial\sigma^i}(\sigma=0,\mu),\qquad i=0,1,2,\ldots,\]
we find at zeroth order ($\sigma^0$):
\bea
K &=& \frac{3\tilde{\alpha}}{2M} \nonumber \\
&=& \frac{3}{M\sqrt{16+j^2(\mu^2-1)+16A_1-16(1-\mu^ 2)(A_0')^2}}, \nonumber
\eea
i.e., a connection between $A_0'=dA_0/d\mu$ and $A_1$. With the help of this relation, we move to order $\sigma^1$ and find an expression for $A_2$ in terms of $A_0$:
\bea
0 &=& 8K^2M^2(1+A_2)-2j^2K^2M^2 + 2\mu\left[ 9-j^2K^2M^2(1-\mu^2)\right]A_0' \nonumber \\
&+& 32K^2M^2(1-\mu^2)A_0'^3 +(1-\mu^2)\left[ 9+32A_0'^2K^2M^2(1-\mu^2)\right]A_0''. \nonumber
\eea
Going further to order $\sigma^2$, we eliminate now $A_1$ and $A_2$ and obtain the relation:
\begin{eqnarray}
0 &=& 243 + 128K^4M^4\left(A_3 +2- j^2 \right)+ 54K^2M^2j^2(1-\mu^2)   \nonumber \\
&+& 3K^4M^4j^4(1-\mu^2)^2 + 768K^4M^4(1-\mu^2)(1-5\mu^2)A_0'^4\nonumber \\
&+& 96K^2M^2(1-3\mu^2)\left[ 9 + K^2M^2j^2(1-\mu^2)\right]A_0'^2\nonumber \\
&-&48K^2M^2(1-\mu^2)^2\left[ 21+64K^2M^2A_0'^2(1-\mu^2)\right]A_0''^2 \nonumber \\
&+& 192K^2M^2A_0'\mu(1-\mu^2)\times\nonumber \\
&&\qquad\times\left[ 27 + K^2M^2( j^2+48A_0'^2)(1-\mu^2)\right]A_0''  \nonumber \\
&-& 32K^2M^2A_0'(1-\mu^2)^2\left[ 27+32K^2M^2A_0'^2(1-\mu^2)\right]A_0'''.\nonumber
\end{eqnarray}

One might get the impression that this sequence could be extended ad infinitum, thus providing at $i$-th order the function $A_{i+1}$ in terms of $A_0$ (by recursively inserting the relations identified before). If this were true, then {\em any} regular function $A_0$ could be prescribed initially, from which all $A_i$ would follow and hence the analytic continuation of $A=A(\sigma,\mu)$ into a vicinity of \scri. \footnote{An open question would be whether, for specifically prescribed $A_0$, this sequence actually converges for all $\sigma\in[0,\sh]$.} 

However, the fact of the matter is that at third order this sequence breaks down since the coefficient in front of $A_4$ vanishes identically. Instead, the third order yields an ordinary differential equation for $A_0$,
\[
0=(1-\mu^2)^2 A_0''''  -  8\mu (1-\mu^2)A_0''' - 4 (1-3\mu^2)A_0'',
\]
which implies
\[A_0''(\mu)=\frac{C_1+C_2\mu}{(1-\mu^2)^2},   \quad\mu=\cos\theta\in[-1,1]\]
with $C_1$ and $C_2$ being constants of integration.
Hence, {\em regular} future null infinity boundary values of $A$ are of the form
\[A_0=c_0+c_1\mu\]
(obtained for $C_1=0=C_2$), which means that they cannot be prescribed arbitrarily. In the following, we restrict ourselves to the 
trivial solution 
\begin{equation}
\label{eq:A0}   A_0(\mu)=0,
\end{equation}
by which we retain coordinates in which the metric coefficients are equatorially symmetric (we also simply neglect an irrelevant constant time shift by putting $c_0=0$). For this choice, the functions $A_1, A_2$ and $A_3$ turn out to be
\begin{eqnarray}
\label{eq:A1}  A_1(\mu)&=& -1+\frac{9}{16K^2M^2}+\frac{j^2}{16}(1-\mu^2)\\
\label{eq:A2}  A_2(\mu)&=&-1 + \frac{j^2}{4}\\
\label{eq:A3}  A_3(\mu)&=&-2-\frac{243}{128K^4M^4}+j^2\left(1-\frac{27(1-\mu^2)}{64K^2M^2}\right) -\frac{3j^4}{128}(1-\mu^2)^2.
\end{eqnarray}
Now, at third order, the equation gives identically zero at both sides (as expected), whereas at all higher orders $i\ge 4$ a relation involving the functions $\{A_4,\ldots,A_{i+1}\}$ emerges. We can argue in the same manner as before. If we prescribe $A_4$ then all $A_i$ for $i>4$ follow and hence the analytic continuation of $A=A(\sigma,\mu)$ into a vicinity of \scri. This time, however, the sequence does not cease at some finite $i$, as the coefficient at $i$-th order in front of $A_{i+1}$ never vanishes. 

Alternatively to prescribing $A_4$ we may require some inner boundary condition which we formulate, for simplicity, in terms of a Dirichlet condition at the event horizon $\sigma=\sh$,
\[A(\sigma=\sh,\mu)=A_{\rm h}(\mu).\]

It is interesting to look at the static, nonrotating situation $j=0$. From the well-known spherically symmetric CMC-slices for the Schwarzschild-solution \cite{brill:2789,Malec:2009} we can read off the explicit solution $A(\sigma,\mu)=A_S(\sigma)$, 

\beq
\label{eq:A_Schwarzschild}
A_S(\sigma)=\int_0^\sigma\left[\frac{f(\tilde\sigma)+a(\tilde\sigma)}{2\tilde\sigma^2(\tilde\sigma-1)f(\tilde\sigma)}
+\frac{1}{\tilde\sigma^2}+\frac{1}{\tilde\sigma}\right]
d\tilde\sigma
\eeq
with\footnote{The auxiliary function $a(\sigma)$ (where we follow the conventional notation) is not to be confused with the Kerr parameter $a=J/M$.}
\beq
f(\sigma)=\sqrt{1-\sigma +a^2(\sigma)},\quad a(\sigma)=\frac{2KM}{3\sigma}-\frac{C\sigma^2}{4M^2}.
\eeq

Slices extending from the horizon up to \scri are obtained for
\[
C>\frac{8}{3}KM^3.
\]
Note that despite the apparently singular structure, $A_{S}(\sigma)$ is analytic for all $\sigma\in[0,1]$, possessing at \scri the Taylor expansion $A_{S}(\sigma)=\sum\limits_{i=0}^\infty A_i\sigma^i$ with the first terms given by (\ref{eq:A0} -- \ref{eq:A3}) for $j=0$.\footnote{In the Schwarzschild case $j=0$, the horizon is located at $\sigma=\sh=1$, cf.~(\ref{eq:sh}).}

\begin{figure}[!ht]
\begin{center}
\includegraphics[clip]{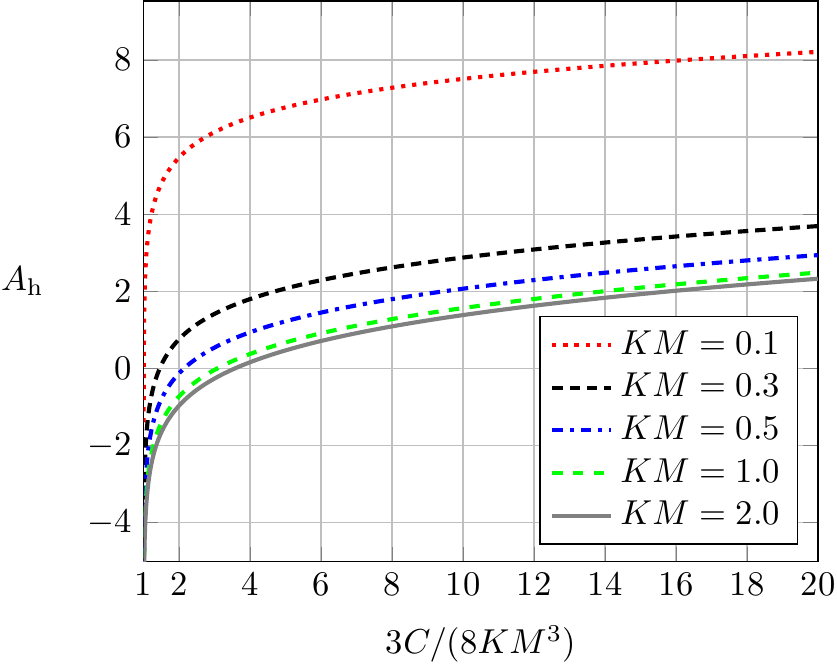}
\caption{The plot shows the relation between the mean curvature $K$, the free parameter $C$ and the boundary value $A_{\rm h}=A_S(\sigma=1)$ of the transformation function $A_S$ in the spherically symmetric Schwarzschild space-time, cf.~(\ref{eq:A_Schwarzschild}). For prescribed $KM$ there is a monotonic relation between $A_{\rm h}$ and $CM^{-2}$. 
}
\label{fig:Ah_of_C}
\end{center}
\end{figure}

We observe that $A_{S}$ depends on $K$ as well as on an additional parameter $C$ which can be related to the boundary value $A_{S}(\sigma=1)=A_{\rm h}$, see figure \ref{fig:Ah_of_C}. Abandoning spherical symmetry and returning to axisymmetry, this free parameter $C$ is replaced by a free function which is encoded in $A_{\rm h}(\mu)$. This being the case, we are able to construct CMC-slices of the Schwarzschild solutions which are not spherically symmetric by simply choosing $A_{\rm h}$ to depend on $\mu$. An example will be discussed below.

\section{The numerical scheme}\label{sec:NumScheme}

We solve equation (\ref{eq:K_explicit}) with a single-domain pseudo-spectral Gauss-Lobatto method. The domain contains all $(\sigma,\mu)\in[0,\sh]\times[-1,1]$. At the exterior boundary $\sigma=0$, we require that $A$ vanishes (cf.~(\ref{eq:A0})) while at $\sigma=\sh$ we impose the Dirichlet boundary condition $A(\sh,\mu)=A_{\rm h}(\mu)$ for some prescribed function $A_{\rm h}$.

For given numerical resolutions $n_\sigma$ and $n_\mu$, the collocation points are located at
\begin{equation*}
\label{eq:Gauss_Lobatto}
\begin{array}{ccll}
\sigma_i &=& \sh\sin^2\left(\dfrac{\pi i}{2(n_\sigma-1)}\right)& (i=0,\ldots, n_\sigma-1)\\[5mm]
\mu_k &=& -\cos\left(\dfrac{\pi k}{n_\mu-1}\right)&(k=0,\ldots, n_\mu-1)\,.
\end{array}
\end{equation*}
The collocation points are the abscissa values of the local extrema of the Chebyshev polynomials. Approximate values for $A(\sigma_i,\mu_k)$ are stored in a vector $\mathbf{f}^{(n_\sigma,n_\mu)}$ (see e.g. \cite{meinel2008relativistic} for a detailed description of the method). From this vector the Chebyshev coefficients as well as the first and second derivatives at the collocation points are approximated. Equation (\ref{eq:K_explicit}) is then solved together with the aforementioned boundary conditions by means of a Newton-Raphson scheme, which uses the ``bi-conjugate gradient stabilized method'' \cite{Barrett:1993} for inverting the Jacobian. This method needs an appropriate preconditioner, for which we utilize a finite difference representation of the Jacobian and invert it with the help of a band diagonal matrix decomposition algorithm (see, e.g. \cite{Press2007numericalrecipes} and references therein). For sufficiently small $j$, we take the known spherically-symmetric solution in the Schwarzschild case (\ref{eq:A_Schwarzschild}) as the initial guess for the Newton-Raphson scheme. We gradually explore the entire regime $j\in[0,1]$ by increasing $j$. Any solution computed serves as the Newton-Raphson guess for the subsequent solution. Note that we encounter no obstacle in attaining the extreme limit $j=1$.
\section{Numerical results}\label{sec:Num_results}
\subsection{Constant mean curvature slices in the Kerr space-time}\label{sec:CMC_Kerr}
\begin{figure}[b!]
\begin{center}
\includegraphics[width=6.cm,clip]{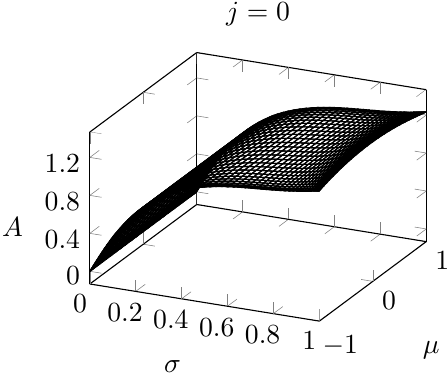}
\includegraphics[width=6.cm,clip]{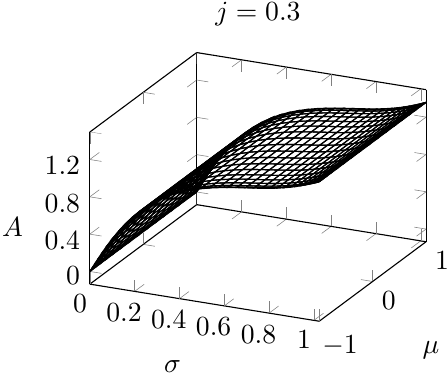}
\includegraphics[width=6.cm,clip]{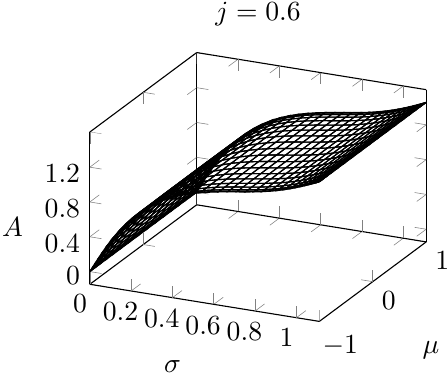}
\includegraphics[width=6.cm,clip]{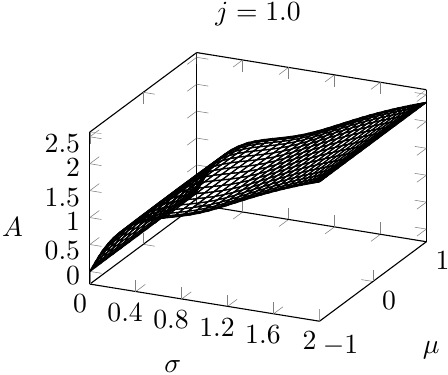}
\caption{The transformation function $A(\sigma,\mu)$ for Kerr Black Holes within the domains $[0,\sigma_h]\times[-1,1]$, with $j\in\{0; 0.3; 0.6; 1\}$ (and corresponding $\sigma_h \in \{1;1.02;1.11;2\}$). The results show a mild dependence on $\mu$ even for highly spinning black holes. }
\label{fig:A_Kerr}
\end{center}
\end{figure}

The construction of transformation functions $A$ requires the prescription of Dirichlet boundary values $A_{\rm h}(\mu)=A(\sh,\mu)$. For simplicity, we choose $A_{\rm h}$ to be a constant, specifically the value $A_S(\sh)$ with $A_S$ and $\sh$ given by (\ref{eq:A_Schwarzschild}) and (\ref{eq:sh}) respectively. This value depends on the two parameters $K$ and $C$ (see figure \ref{fig:Ah_of_C}), and we adopt a numerically favorable choice, namely 
\beq
K\approx 0.33 M^{-1},\qquad C\approx 2.88M^2,\label{eq:C_and_K}
\eeq
which has been identified in \cite{Schinkel:2013zm} where we aimed at a rapidly converging spectral expansion of the conformal lapse. 

In figure \ref{fig:A_Kerr} the transformation function $A$ is shown for $j\in\{0; 0.3; 0.6; 1\}$, i.e.~the last computation refers to the extremal case. We notice that $A$ depends only weakly on $\mu$, even for highly spinning Kerr black holes. In order to reach extreme limit, only three intermediate solutions (for $j\in\{0.25; 0.5; 0.75\}$) need to be computed.

\begin{figure}[ht]
\begin{center}
\includegraphics[clip]{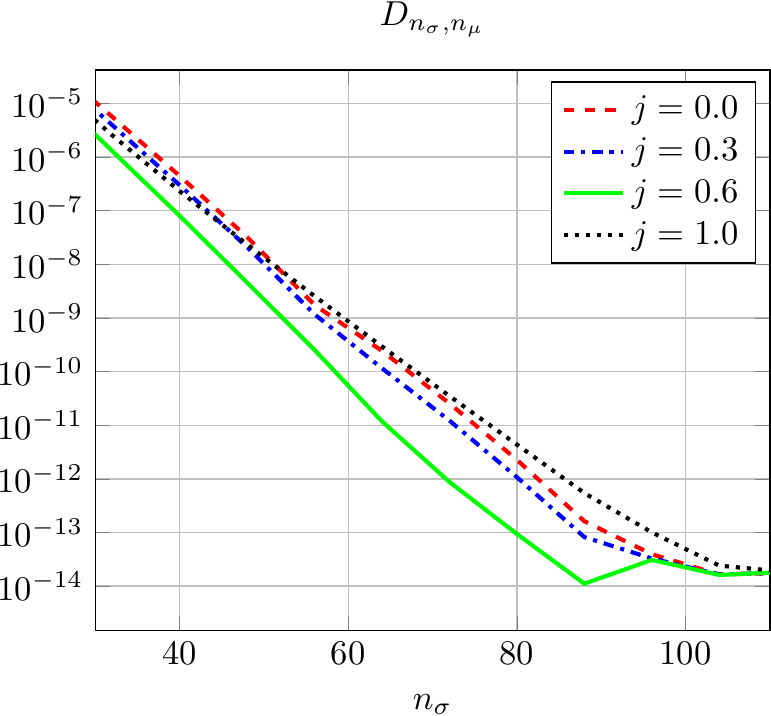}
\caption{The maximal deviation $D_{n_{\sigma},n_\mu}$ (cf.~(\ref{eq:maximal_deviation})) is plotted against the spectral resolution $n_\sigma$ (here $n_\sigma=4n_\mu$) for different specific angular momentum $j$. The exponential decay is a good indication for the regularity of the solution.}
\label{fig:convergence_Kerr}
\end{center}
\end{figure}

In figure \ref{fig:convergence_Kerr}, we plot the quantity
\beq
D_{n_{\sigma},n_\mu}=\sup_{\sigma,\mu}|A_{n_{\sigma},n_\mu} -A_{200,50}|,\label{eq:maximal_deviation}
\eeq
which describes the maximal deviation of $A(\sigma,\mu)$ with respect to a numerical reference solution with resolution $(n_{\sigma},n_\mu)=(200,50)$. The exponential decay of $D_{n_{\sigma},n_\mu}$ is a good indication for the regularity of the solution.
\subsection{Constant mean curvature slices in the Schwarzschild space-time}\label{sec:CMC_Schwarzschild}
In this section we demonstrate that CMC-slices in the Schwarzschild solution need not necessarily be the well-known spherically symmetric foliations given through (\ref{eq:A_Schwarzschild}). As discussed in \autoref{sec:ConstructionCMC}, slices which are not spherically symmetric emerge if the boundary function $A_{\rm h}(\mu)=A(1,\mu)$ is chosen to depend on $\mu$. Here we consider the specific example
\beq
A_{\rm h}(\mu) = A_S(1) + \epsilon\cos(\mu)\label{eq:CMC_Schwarzschild_innerBoundary},
\eeq
where $A_S(1)$ is the horizon value in the spherically symmetric case (cf.~(\ref{eq:A_Schwarzschild})). 
The pair $(K,C)$ is chosen as before (cf.~(\ref{eq:C_and_K})), and we thus obtain $A_S(1)\approx{1.33}$. Starting from the spherically symmetric situation $\epsilon=0$ we gradually increase $\epsilon$ up to $\epsilon=1/2$, taking any solution computed as the Newton-Raphson guess for the subsequent solution (see \autoref{sec:NumScheme}).

\begin{figure}[!h]
\begin{center}
\includegraphics[width=5.5cm,clip]{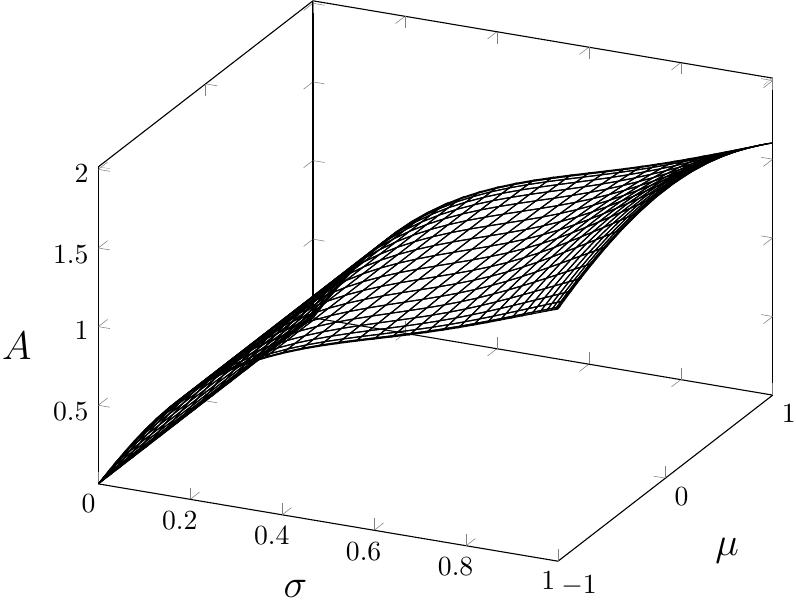}
\includegraphics[width=4.8cm,clip]{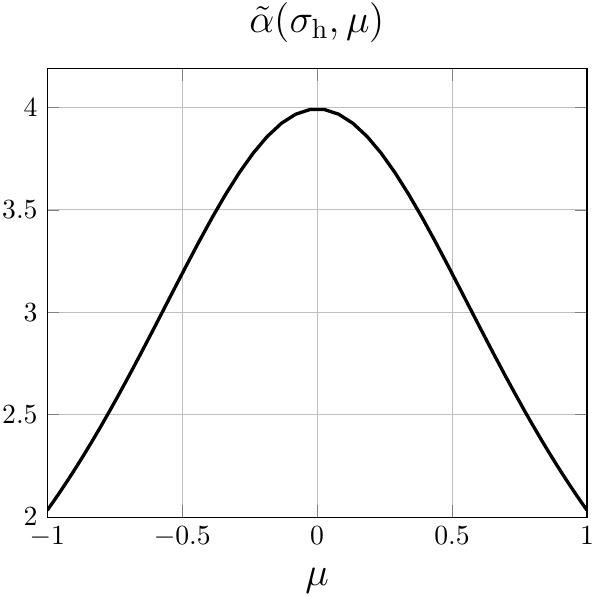}
\includegraphics[width=5.1cm,clip]{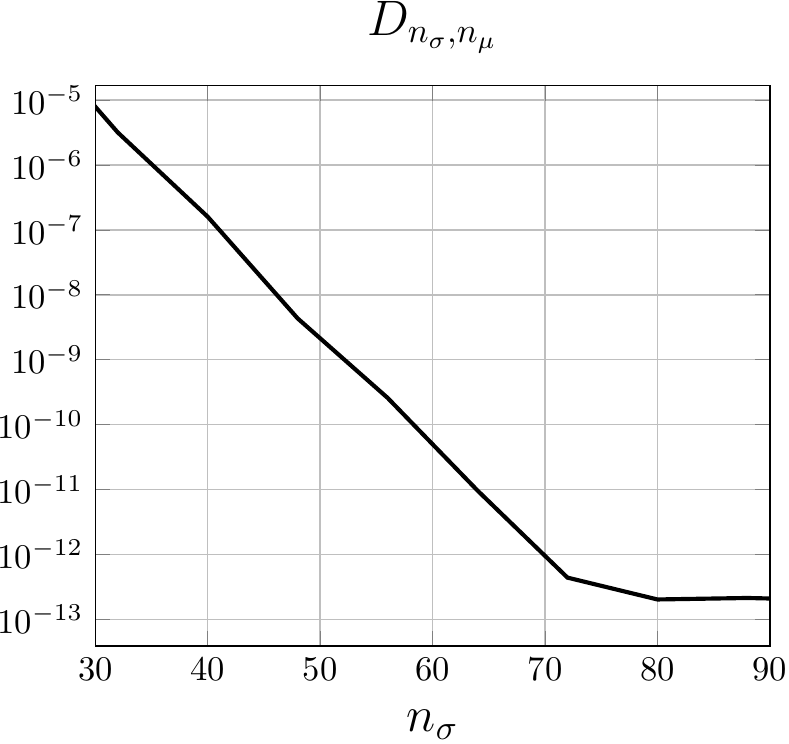}
\caption{For the Dirichlet boundary data (\ref{eq:CMC_Schwarzschild_innerBoundary}) with $\epsilon=1/2$, the transformation function $A=A(\sigma,\mu)$ is shown (left panel) as well as the conformal lapse $\tilde\alpha(1,\mu)$
at the horizon (middle panel). In the right panel, the maximal deviation $D_{n_{\sigma},n_\mu}$ (cf.~(\ref{eq:maximal_deviation})) is plotted against the spectral resolution $n_\sigma$ (here $n_\sigma=4n_\mu$).}
\label{fig:Schwarzschild_nonsymmetric}
\end{center}
\end{figure}

For $\epsilon=1/2$, figure \ref{fig:Schwarzschild_nonsymmetric} shows the resulting $A(\sigma,\mu)$ as well as the conformal lapse $\tilde\alpha(1,\mu)$ at the inner boundary (here the $\mu$-dependence becomes apparent).
In all calculations we find a rapid exponential decay of the Chebyshev coefficients of $A(\sigma,\mu)$. Also in figure \ref{fig:Schwarzschild_nonsymmetric}, we plot the deviations $D_{n_{\sigma},n_\mu}$ (cf.~\ref{eq:maximal_deviation})
where again a numerical reference solution with resolution $(n_{\sigma},n_\mu)=(200,50)$ was taken. The exponential decay of $D_{n_{\sigma},n_\mu}$ is once more a good indication for the regularity of the solution.
\section{Discussion}\label{sec:Discussion}

In this paper we constructed constant mean curvature slices within the Kerr space-time. We found that for given mean curvature value $K$, there exists a class of CMC-slices which are characterized by a free boundary function $A_{\rm h}=A_{\rm h}(\cos\theta)$. Example calculations were performed for the specific choice of constant boundary data, $A_{\rm h}={\rm constant.}$ 
The pseudo-spectral scheme yields exponentially converging solutions, thus demonstrating the regularity of the slices, which applies to the entire parameter realm $j\in[0,1]$ of Kerr black holes. As a by-product, we constructed CMC-slices in the Schwarzschild solution which are not spherically symmetric. 

For a future direction, the CMC-slices of the Kerr space-time can be taken as a starting point for the computation of initial data for perturbed Kerr black holes on hyperboloidal CMC-slices. One way to accomplish this goal would be to adopt the methods described in \cite{Schinkel:2013zm}, which work on non-conformally flat slices with the entirety of all constraints and thus yield  arbitrarily small perturbations of the Kerr metric, even very close to the extremal case.
\section*{Acknowledgments}
We are very greateful to  Niall Ó Murchadha for bringing this topic to our attention. Also, it is a pleasure to thank L.~Buchman and J.~L.~Jaramillo for many valuable discussions and N. Johnson-McDaniel for carefully reading the manuscript. This work was supported by the DFG-grant SFB/Transregio 7 ``Gravitational Wave Astronomy''.
\section*{Bibliography}
\bibliography{bibitems}

\begin{thebibliography}{10}

\bibitem{Bondi1962}
H.~Bondi, M.G.J. van~der Burg, and A.W.K. Metzner.
\newblock {Gravitational waves in general relativity. 7. Waves from
  axisymmetric isolated systems}.
\newblock {\em Proc.Roy.Soc.Lond.}, A269:21--52, 1962.

\bibitem{Stewart1989}
J.M. Stewart.
\newblock Numerical relativity iii. the {B}ondi mass revisited.
\newblock {\em Proc. R. Soc. London, Ser. A}, 424:211--222, 1989.

\bibitem{Pfeiffer:2012pc}
Harald~P. Pfeiffer.
\newblock {Numerical simulations of compact object binaries}.
\newblock {\em Class.Quant.Grav.}, 29:124004, 2012.

\bibitem{Frauendiener:ConformalInfinity}
Jörg Frauendiener.
\newblock Conformal infinity.
\newblock {\em Living Reviews in Relativity}, 7(1), 2004.

\bibitem{Friedrich:1983}
Helmut Friedrich.
\newblock Cauchy problems for the conformal vacuum field equations in general
  relativity.
\newblock {\em Communications in Mathematical Physics}, 91:445--472, 1983.

\bibitem{Hubner:2000pb}
Peter Hübner.
\newblock {From now to timelike infinity on a finite grid}.
\newblock {\em Class.Quant.Grav.}, 18:1871--1884, 2001.

\bibitem{Zenginoglu:2008pw}
Anıl Zenginoğlu.
\newblock {Hyperboloidal evolution with the Einstein equations}.
\newblock {\em Class.Quant.Grav.}, 25:195025, 2008.

\bibitem{Moncrief2009}
Vincent Moncrief and Oliver Rinne.
\newblock {Regularity of the Einstein equations at future null infinity}.
\newblock {\em Classical and Quantum Gravity}, 26(12):125010, 2009.

\bibitem{Rinne:2009qx}
Oliver Rinne.
\newblock {An Axisymmetric evolution code for the Einstein equations on
  hyperboloidal slices}.
\newblock {\em Class.Quant.Grav.}, 27:035014, 2010.

\bibitem{Bardeen:2011ip}
James~M. Bardeen, Olivier Sarbach, and Luisa~T. Buchman.
\newblock {Tetrad formalism for numerical relativity on conformally
  compactified constant mean curvature hypersurfaces}.
\newblock {\em Phys.Rev.}, D83:104045, 2011.

\bibitem{Bardeen:2011pd}
James~M. Bardeen and Luisa~T. Buchman.
\newblock {Bondi-Sachs Energy-Momentum for the CMC Initial Value Problem}.
\newblock {\em Phys.Rev.}, D85:064035, 2012.

\bibitem{Andersson:1992yk}
Lars Andersson, Piotr~T. Chruściel, and Helmut Friedrich.
\newblock {On the Regularity of solutions to the Yamabe equation and the
  existence of smooth hyperboloidal initial data for Einsteins field
  equations}.
\newblock {\em Commun.Math.Phys.}, 149:587--612, 1992.

\bibitem{Andersson:1993we}
Lars Andersson and Piotr~T. Chruściel.
\newblock {Hyperboloidal Cauchy data for vacuum Einstein equations and
  obstructions to smoothness of null infinity}.
\newblock {\em Phys. Rev. Lett.}, 70:2829--2832, May 1993.

\bibitem{Andersson1994}
Lars Andersson and Piotr~T. Chruściel.
\newblock {On “hyperboloidal” Cauchy data for vacuum einstein equations and
  obstructions to smoothness of Scri}.
\newblock {\em Communications in Mathematical Physics}, 161:533--568, 1994.

\bibitem{Frauendiener:1998ud}
Jörg Frauendiener.
\newblock {Calculating initial data for the conformal Einstein equations by
  pseudo-spectral methods}.
\newblock {\em Journal of Computational and Applied Mathematics},
  109(1–2):475 -- 491, 1998.

\bibitem{Husa:2005ns}
Sascha Husa, Carsten Schneemann, Tilman Vogel, and Anil Zenginoğlu.
\newblock {Hyperboloidal data and evolution}.
\newblock {\em AIP Conf.Proc.}, 841:306--313, 2006.

\bibitem{Bowen:1980yu}
Jeffrey~M. Bowen and Jr. York, James~W.
\newblock {Time asymmetric initial data for black holes and black hole
  collisions}.
\newblock {\em Phys.Rev.}, D21:2047--2056, 1980.

\bibitem{Buchman:2009ew}
Luisa~T. Buchman, Harald~P. Pfeiffer, and James~M. Bardeen.
\newblock {Black hole initial data on hyperboloidal slices}.
\newblock {\em Phys.Rev.}, D80:084024, 2009.

\bibitem{Lovelace2008}
Geoffrey Lovelace, Robert Owen, Harald~P. Pfeiffer, and Tony Chu.
\newblock Binary-black-hole initial data with nearly extremal spins.
\newblock {\em Phys. Rev. D}, 78:084017, Oct 2008.

\bibitem{Garat:2000pn}
Alcides Garat and Richard~H. Price.
\newblock {Nonexistence of conformally flat slices of the Kerr space-time}.
\newblock {\em Phys.Rev.}, D61:124011, 2000.

\bibitem{brill:2789}
Dieter~R. Brill, John~M. Cavallo, and James~A. Isenberg.
\newblock {K-surfaces in the Schwarzschild space-time and the construction of
  lattice cosmologies}.
\newblock {\em Journal of Mathematical Physics}, 21(12):2789--2796, 1980.

\bibitem{Malec:2009}
Edward Malec and Niall \'O~Murchadha.
\newblock {General spherically symmetric constant mean curvature foliations of
  the Schwarzschild solution}.
\newblock {\em Phys. Rev. D}, 80:024017, Jul 2009.

\bibitem{Gentle:2000aq}
Adrian~P. Gentle, Daniel~E. Holz, Arkady Kheyfets, Pablo Laguna, Warner~A.
  Miller, et~al.
\newblock {Constant crunch coordinates for black hole simulations}.
\newblock {\em Phys.Rev.}, D63:064024, 2001.

\bibitem{Zenginoglu:2008wc}
Anıl Zenginoğlu.
\newblock {A Hyperboloidal study of tail decay rates for scalar and Yang-Mills
  fields}.
\newblock {\em Class.Quant.Grav.}, 25:175013, 2008.

\bibitem{Zenginoglu:2008uc}
Anıl Zenginoğlu, Darío Núñez, and Sascha Husa.
\newblock {Gravitational perturbations of Schwarzschild spacetime at null
  infinity and the hyperboloidal initial value problem}.
\newblock {\em Class.Quant.Grav.}, 26:035009, 2009.

\bibitem{Tuite2013}
Patrick Tuite and Niall \'O~Murchadha.
\newblock Constant mean curvature slices of the reissner-nordström spacetime.
\newblock July 2013.

\bibitem{Racz:2011qu}
István Rácz and Gábor~Zsolt Tóth.
\newblock {Numerical investigation of the late-time Kerr tails}.
\newblock {\em Class.Quant.Grav.}, 28:195003, 2011.

\bibitem{Harms:2013ib}
Enno Harms, Sebastiano Bernuzzi, and Bernd Brügmann.
\newblock {Numerical solution of the 2+1 Teukolsky equation on a hyperboloidal
  and horizon penetrating foliation of Kerr and application to late-time
  decays}.
\newblock {\em Class.Quant.Grav.}, 30:115013, 2013.

\bibitem{Thornburg:1987}
Jonathan Thornburg.
\newblock Coordinates and boundary conditions for the general relativistic
  initial data problem.
\newblock {\em Classical and Quantum Gravity}, 4(5):1119, 1987.

\bibitem{Seidel:1992}
Edward Seidel and Wai-Mo Suen.
\newblock Towards a singularity-proof scheme in numerical relativity.
\newblock {\em Phys. Rev. Lett.}, 69:1845--1848, Sep 1992.

\bibitem{baumgarte2010numerical}
T.W. Baumgarte and S.L. Shapiro.
\newblock {\em Numerical Relativity: Solving Einstein's Equations on the
  Computer}.
\newblock Cambridge University Press, 2010.

\bibitem{Schinkel:2013zm}
David Schinkel, Rodrigo~Panosso Macedo, and Marcus Ansorg.
\newblock {Inital data for perturbed Kerr black holes on hyperboloidal slices}.
\newblock 2013.

\bibitem{Zenginoglu:2007jw}
Anıl Zenginoğlu.
\newblock {Hyperboloidal foliations and scri-fixing}.
\newblock {\em Class.Quant.Grav.}, 25:145002, 2008.

\bibitem{Wald:1984}
Robert~M Wald.
\newblock {\em General Relativity}.
\newblock University of Chicago Press, 1984.

\bibitem{MTW:1973}
C.W. Misner, K.S. Thorne, and J.A. Wheeler.
\newblock {\em Gravitation}.
\newblock W.H. Freeman, San Francisco, 1973.

\bibitem{meinel2008relativistic}
Reinhard Meinel, Marcus Ansorg, Andreas Kleinwächter, Gernot Neugebauer, and
  D.~Petroff.
\newblock {\em Relativistic Figures of Equilibrium}.
\newblock Cambridge University Press, 2008.

\bibitem{Barrett:1993}
R.~Barrett, M.~Berry, T.~F. Chan, J.~Demmel, J.~Donato, J.~Dongarra,
  V.~Eijkhout, R.~Pozo, C.~Romine, and H.~Vander Vorst.
\newblock {\em Templates for the Solution of Linear Systems: Building Blocks
  for Iterative Methods, 2nd Edition}.
\newblock SIAM, Philadelphia, PA, 1994.

\bibitem{Press2007numericalrecipes}
William~H. Press, Saul~A. Teukolsky, William~T. Vetterling, and Brian~P.
  Flannery.
\newblock {\em Numerical Recipes 3rd Edition: The Art of Scientific Computing}.
\newblock Cambridge University Press, New York, NY, USA, 3 edition, 2007.

\end{thebibliography}
\end{document}